\documentclass[%
 reprint,
amsmath,amssymb,
aps,
pra,
twocolumn,
]{revtex4-2}
\usepackage{braket}
\usepackage{graphicx}% Include figure files
\usepackage{bm}% bold math
\usepackage{float}
\usepackage{color}
\usepackage{hyperref}% add hypertext capabilities
\usepackage{placeins}
\usepackage[normalem]{ulem}

\begin{document}

\title{Tunneling assisted interference in double-well levitodynamics}
\author{Yue Ma$^1$}
\author{Endre Bokor$^1$}
\author{M. S. Kim$^1$}

\affiliation{$^1$Blackett Laboratory, Imperial College London, London SW7 2AZ, United Kingdom\\
}

\begin{abstract}

Levitated particles trapped in double-well potentials provide a promising platform for exploring quantum dynamics beyond the harmonic regime. We show that such systems exhibit a distinctive quantum signature arising from tunneling-assisted interference. Unlike classical dynamics, where the two wells are disconnected, in the quantum case, quantum tunneling enables eigenstates to delocalize into the other well, creating nonzero overlaps between adjacent eigenstates that introduce low-frequency components into the evolution of the mean particle position. By comparing quantum and classical dynamics of particles in one-dimensional double-well potentials, we demonstrate that this interference is uniquely quantum and can be identified through the oscillation of the average position alone. We further relate the observed spectral peaks to the structure of the energy eigenvalues. The effect provides a promising probe of the quantum signature of levitated particle motion, and has fundamental importance as it predicts the demonstration of two major phenomena of quantum mechanics, quantum tunneling and quantum interference, at the same time and on the same system.

\end{abstract}

%\date{\today}

\maketitle

%\section{Introduction}

%[Double well potential, introduction say all the importance of double wells but not for large quantum effects. Mention double well potential for chemical reaction(?) too, see my interview slides. For everything else the exact shape of the double well doesn’t matter. Except levitodynamics, but people haven’t studied these yet.]

The harmonic oscillator is among the most important models in physics because of its analytical simplicity and broad applicability. Its dynamics, however, are fundamentally linear and therefore unable to capture many nonlinear phenomena. A typically extension is the double-well potential, which has served as a cornerstone model across diverse areas of physics. Double-well systems have been extensively studied in molecular physics, where the splitting of the lowest-energy doublet is associated with quantum tunneling between the wells~\cite{halataei2017tunnel,fiechter2026ring}, as well as in trapped-ion platforms~\cite{retzker2008double}, ultracold atomic gases~\cite{folling2007direct,bonneau2018characterizing}, and chemical reaction dynamics~\cite{pollak2023recent}.

In many of these contexts, the double-well potential primarily serves to define two distinct configurations. Theoretical descriptions are therefore often reduced to a small number of effective parameters such as the barrier width and height, while details of the potential shape play only a secondary role. As a consequence, much of the rich physics associated with the full structure of the double-well potential remains unexplored.

A particularly promising setting in which the detailed form of the double-well potential becomes essential is levitodynamics~\cite{gonzalez2021levitodynamics}. Recent experimental advances have realized double-well trapping potentials for micro- and nanoscale particles using a variety of optical and electromechanical techniques~\cite{rondin2017direct,ricci2017optically,almeida2023trapping,sun2024tunable,dago2024stabilizing,toftul2025hopping,mlynavr2026feedback}. Despite rapid experimental progress, the quantum dynamics of levitated particles in double-well potentials remain comparatively unexplored. Existing theoretical studies have largely focused on limiting cases. These include wide double-well configurations in which the dynamics are effectively restricted to a single well~\cite{roda2024macroscopic}, as well as analyses based on inverted parabolic barriers that capture local properties of the barrier region but not the physics of a bounded double-well potential~\cite{romero2017coherent}.

In this Letter, we show that a particle moving in a one-dimensional double-well potential exhibits a distinctive dynamical quantum signature originating from tunneling-assisted interference between eigenstates localized in opposite wells. Conceptually, it unambiguously connects quantum interference with the original, intuitive interpretation of quantum tunneling as wavefunction exceeding the boundary of a trapping potential, therefore combining these two main quantum phenomena in one system. Practically, it offers multiple experimental advantages. The effect emerges at the beginning of the time evolution without the need of waiting for long-time quantum revivals, and it requires only measurements of the particle mean position without the need of reconstructing the full quasi-probability distribution, thereby providing a clear route toward observing quantum behavior in levitodynamics.

\begin{figure}[t]
\centering
\includegraphics[width=0.48\textwidth]{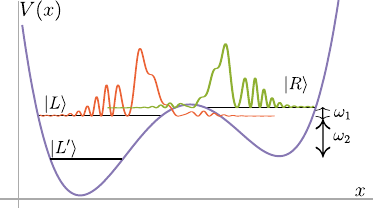}
\caption{The qutrit toy model with toy wavefunctions for the qualitative explanation of tunneling assisted interference in a double-well potential. Quantum tunneling allows the wavefunction $|L(R)\rangle$ to extend to the right (left) well, leading to a non-zero interference amplitude $\langle L|\hat{x}|R\rangle$. This in turn contributes to a low-frequency component $\omega_1$ in the time evolution of the mean value of the particle position, which is unavailable in the classical case.}\label{fig:toy_model}
\end{figure}

\noindent \textit{Toy model.} --- The working principle of the tunneling assisted interference can be straightforwardly understood via a simplified model where only three  eigenstates of a double-well potential are considered (see Fig.~\ref{fig:toy_model}). The eigenstates $|L\rangle$ and $|L'\rangle$ dominantly overlap with the left well, with eigenvalues $E_L$ and $E_{L'}$, respectively. The eigenstate $|R\rangle$, with eigenvalue $E_R$, mostly overlaps with the right well. We define frequencies that are associated with the dynamics as $\omega_1=(E_R-E_L)/\hbar$ and $\omega_2=(E_L-E_{L'})/\hbar$.

For an initial state as the superposition of these eigenstates, $|\psi_i\rangle=\alpha_0|L\rangle+\beta_0|R\rangle+\gamma_0|L'\rangle$, the time evolution of the mean value of the position is
\begin{align}
    &\langle \hat{x}(t)\rangle\nonumber\\
    =&|\alpha_0|^2\langle L|\hat{x}|L\rangle+|\beta_0|^2\langle R|\hat{x}|R\rangle+|\gamma_0|^2\langle L'|\hat{x}|L'\rangle\nonumber\\
    +&2\alpha_0\beta_0\langle L|\hat{x}|R\rangle \cos\omega_1t+2\alpha_0\gamma_0\langle L|\hat{x}|L'\rangle \cos\omega_2t\nonumber\\
    +&2\beta_0\gamma_0\langle R|\hat{x}|L'\rangle \cos(\omega_1+\omega_2)t.
\end{align}
For most double-well potentials that we will consider,
%(except the case of near-degeneracy as discussed in \textcolor{red}{ the accompanying long paper} and Ref.~\cite{ma2020quantum})
the dominant low-frequency contributions to the dynamics of $\langle \hat{x}(t)\rangle$ come from $\omega_1$ and $\omega_2$, which are of the same order of magnitude and approximately $\omega_1\approx\omega_2/2$. However, the $\omega_1$ frequency component only shows up in the time evolution of $\langle \hat{x}(t)\rangle$ if the amplitude $\langle L|\hat{x}|R\rangle$ is nonzero. This is the most important feature of the quantum tunneling assisted interference. If the system is classical instead, no tunneling is allowed, meaning that $\langle L|\hat{x}|R\rangle=0$ is strictly set. The observation of the frequency $\omega_1$ in $\langle \hat{x}(t)\rangle$ therefore reveals the quantum effect.

%Leading order approximation for the actual system...

\begin{figure}[t]
\centering
\includegraphics[width=0.48\textwidth]{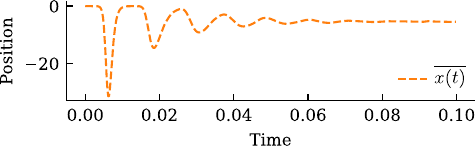}
\caption{The dimensionless classical average dynamics of a particle trapped in a double-well potential, where the average is taken over the ensemble of initial conditions matching the quantum fluctuation of the vacuum state. Real experimental parameters from Ref.~\cite{rondin2017direct} are used. It is straightforward to map back to the dimensional dynamics as described in the main text.}\label{fig:largeC}
\end{figure}

\noindent \textit{The setup and the dimensionless transformation.} --- We consider the one-dimensional dynamics of a particle trapped in a double-well potential, whose Hamiltonian is $H_r(x_r,p_r)=p_r^2/2m+A_rx_r^4+B_rx_r^3+C_rx_r^2+D_rx_r+E_r$, where $p_r$ is the momentum of the particle, $m$ is the mass of the particle, and $x_r$ is the position of the particle. Without loss of generality, we can set $D_r=0$ by defining the axis origin $x_r=0$ as the position of the local maximum of the double-well potential, and set $E_r=0$ as it is a constant shift of energy. For simplicity, we transform the dynamics to a dimensionless one by defining $x=\alpha x_r$ and $t=\gamma t_r$, with $\alpha=(A_rm)^{1/6}\hbar^{-1/3}$ and $\gamma=(A_r\hbar)^{1/3}m^{-2/3}$. Here $\hbar=1.055\times10^{-34}\ \mathrm{kg\cdot m^2\cdot s^{-1}}$ is the reduced Planck's constant, and note that while $x_r$ and $t_r$ have the physical units of $\mathrm{m}$ and $\mathrm{s}$, respectively,  after the transformation $x$ and $t$ are dimensionless. The dimensionless Hamiltonian is thus
\begin{equation}\label{eq:DimLessHam}
    H=\frac{1}{2}p^2+x^4+Bx^3+Cx^2.
\end{equation}
For an initial state centered near $x=0$ (the local maximum of the potential), as argued in the previous section, we expect the quantum average position $\langle \hat{x}(t) \rangle$ to show the interference effect of having a low frequency component about half of the classical counterpart. In Fig.~\ref{fig:largeC}, we show the classical average of the position dynamics, with parameters of particle mass and potential depths taken from the ones reported from an early experimental work~\cite{rondin2017direct}: The particle is a silica sphere with radius $68\ \mathrm{nm}$, the local maximum of the double-well potential is characterized as an inverse parabola with angular frequency $2\pi\times51\ \mathrm{kHz}$, and the depths of the two local minima with respect to the local maximum are $4k_BT$ and $5k_BT$, where the temperature is $T=300\ \mathrm{K}$ and $k_B$ is the Boltzmann constant. Here the classical average is taken as the ensemble average over an initial phase space distribution matching the vacuum state. The dynamics map back to the physical dimensions with parameters $\alpha\approx2\times10^9m^{-1}$, $\gamma\approx150 s^{-1}$, noting that the rescaling between the dimensionless dynamics and the dimensional one does not alter the initial state itself~\cite{ma2023eliminating}. Simulating the quantum dynamics is beyond reach considering the large quantum number associated with the real experimental parameters, therefore in the following sections we will consider smaller systems. However, we may estimate that, as the quantum effect features a lower frequency and a larger and more persistent amplitude~\cite{ma2020quantum} than the classical case, the resolutions in time and in distance required for observing the quantum effect are within experimental reach. Moreover, reducing the size of the particle and making the features of the trapping potential sharper will bring in further improvements. 

%\noindent \textit{Setup, dimensionless transformation and classical average} --- We expect the quantum oscillations (interference pattern) to have a low frequency component about half of the classical frequency, which after bringing back the dimensions happens reasonably fast. We also expect persistent quantum oscillations because of the narrow peaks in the quantum case (see later and cite my rotation papers).

%Bringing back the dimensions by rescaling do not change the quantum state (vacuum state). Cite my nonlinear optics paper.

\begin{figure}[t]
\centering
\includegraphics[width=0.48\textwidth]{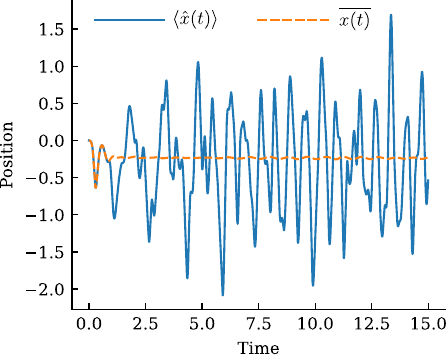}
\caption{The quantum (blue solid line) and classical (orange dashed line) time evolution of the average position of a particle trapped in a double-well potential with dimensionless parameters $B=2$ and $C=-100$. The quantum average refers to the vacuum state as the initial state, while the classical average is taken over the ensemble of initial conditions matching the quantum fluctuation of the vacuum state.
%Quantum v.s. classical time evolution of mean position value. Mention the size of the double-well itself with $B=2$, $C=-100$: the minima of the two wells are around $\pm7$. Looking at the quantum minimal or maximal to read out the low frequency $\omega_1$ component.
}\label{fig:QCtime}
\end{figure}

\noindent \textit{Time evolution of the average position.} --- In order to demonstrate the strong quantum signature contained in the time evolution of the average position, we simulate both the quantum and classical dynamics governed by Eq.~\eqref{eq:DimLessHam}, with dimensionless parameters $B=2$, $C=-100$, corresponding to two local minima around $x=\pm7$ and one local maximum at $x=0$. The initial conditions are the same as those chosen for Fig.~\ref{fig:largeC}. The results are shown in Fig.~\ref{fig:QCtime}. The quantum average position $\langle \hat{x}(t) \rangle$ differs significantly from the classical average position $\overline{x(t)}$. By observing the maximal and minimal of the quantum average, we can resolve a frequency (labeled as $\omega_1$ in our toy model) that is much lower than the classical central frequency (labeled as $\omega_2$ in our toy model) -- the effect of tunneling assisted interference. Additionally, the oscillations in the quantum case persist much longer than those in the classical case. This is due to the different widths of the frequency distributions, resulting from the occupation of eigenstates in the quantum case and of phase space in the classical case. It is thus beyond our toy model, but similar analyses have been made in Ref.~\cite{ma2020quantum}.
%and \textcolor{red}{the accompanying long paper}. 

The quantum-tunneling induced oscillations, as shown in Fig.~\ref{fig:QCtime}, deviate from the classical prediction in a very strong way. It happens at the beginning of the time evolution, when the state just starts to spread, and the mean value of the position already reveals the quantum oscillation fringes. This is in stark contrast with systems featuring harmonic trap and weak nonlinearity, such as those based on Kerr effect, where quantum effects require either long time revivals~\cite{ma2020optical} or the reconstruction of the full state statistics such as the Husimi-Q function~\cite{ma2022unifying,milburn1986quantum}.

%[in later sections] Mention Kerr effect and here no need to construct Q function. Cite my squeezing paper and Milburn gate paper.

\begin{figure}[t]
\centering
\includegraphics[width=0.48\textwidth]{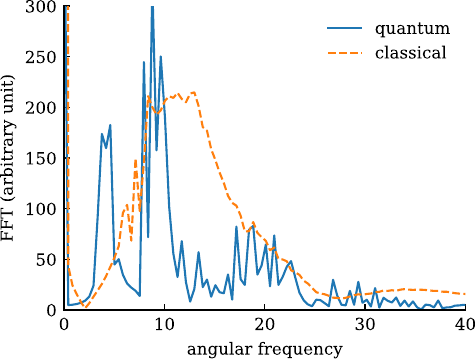}
\caption{Fast Fourier transform of the quantum (blue solid line) and classical (orange dashed line) dynamics of the average position as shown in Fig.~\ref{fig:QCtime}, where $B=2$ and $C=-100$. While the classical case has one wide peak, in the quantum case, this peak is much narrower, and there is an additional low-frequency peak. The existence of zero-frequency peaks for both cases is because the position does not oscillation around zero but instead has an offset. It is irrelevant to our consideration of tunneling-assisted interference effect.
%Rescaled due to the lowest frequency peak.
}\label{fig:FFT}
\end{figure}

\noindent \textit{Spectral peaks.} --- We will next interpret the tunneling assisted interference more quantitatively. Fig.~\ref{fig:FFT} shows the fast Fourier transform of the results of $\langle \hat{x}(t) \rangle$ and $\overline{x(t)}$ plotted in Fig.~\ref{fig:QCtime}. For non-zero frequencies, the quantum spectrum features two main peaks, centered at $\omega_1\approx4.4$ and $\omega_2\approx10$. In contrast, the classical spectrum features only one peak at $\omega_2\approx10$, but much wider than the quantum case.

\begin{figure}[t]
\centering
\includegraphics[width=0.48\textwidth]{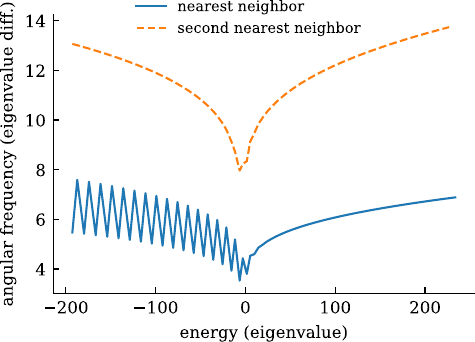}
\caption{The quantum frequency-energy relations based on the eigenvalues of the Hamiltonian Eq.~\eqref{eq:DimLessHam} with $B=2$ and $C=-100$, including both the nearest neighbor frequency branch (blue solid line) and the second nearest neighbor frequency branch (orange dashed line). See the main text for the precise definitions. }\label{fig:f-E}
\end{figure}

\noindent \textit{Frequency-energy relation} --- Now we will give a further interpretation of the two dominant quantum spectral peaks. The dynamics of a quantum system is fundamentally determined by its Hamiltonian. We have thus numerically diagonalized the Hamiltonian Eq.~\eqref{eq:DimLessHam} with parameters $B=2$ and $C=-100$. Straightforwardly from Schr\"odinger's equation, we can define quantum frequencies as $\omega_{i,j}=E_i-E_j$, where $E_{i(j)}$ are eigenvalues of $H$, remembering that in the dimensionless representation $\hbar=1$. The lowest order contributions to the quantum frequencies, namely, from the difference of the nearest-neighboring eigenvalues, and the second-nearest-neighboring eigenvalues, are sufficient to consider for classical-like initial states such as the vacuum state. Once we order the eigenvalues by $E_i>E_j$ for $i>j$, we can define the nearest neighbor frequency-energy curve as the relation between $\omega_{i+1,i}=E_{i+1}-E_i$ and $E_{i+1,i}=(E_{i+1}+E_i)/2$, and the second nearest neighbor frequency-energy curve as the relation between $\omega_{i+2,i}=E_{i+2}-E_i$ and $E_{i+2,i}=(E_{i+2}+E_i)/2$. The results are shown in Fig.~\ref{fig:f-E}. The regions of smaller-than-zero eigenvalues are of special interest, as they manifest tunneling effect that is classically forbidden. Specifically, as they lie below the local maximum of the double-well potential, their corresponding eigenstates expressed in the position basis have dominant support in only one of the two wells, and moreover, the dominant support alternates between the two wells, namely, if $\langle x|E_i\rangle$ is largely in one well,  $\langle x|E_{i-1}\rangle$ and $\langle x|E_{i+1}\rangle$ are both largely in the other well. Importantly, these eigenstates are not strictly bounded to one well. For instance, $\langle x|E_i\rangle$ can be dominantly in the left well but with a small support to the right well, and $\langle x|E_{i\pm1}\rangle$ are dominantly in the right well but with a small support to the left well.
%(\textcolor{red}{see the accompanying long paper}). 
This is the quantum tunneling effect, and it is suppressed as the energy moves away from the potential local maximum, i.e., towards the bottom of the wells. Tunneling facilitates the contribution of the nearest-neighboring frequencies to the dynamics of $\langle \hat{x}(t) \rangle$, while the second-nearest-neighboring frequencies contribute in both the quantum and classical cases, as we have qualitatively explained in the toy model. Indeed, from our numerical simulation, the spectral peak $\omega_1\approx4.4$, which is uniquely quantum, sits within the nearest neighbor frequency-energy curve in Fig.~\ref{fig:f-E}. The other spectral peak $\omega_2\approx10$, which is featured in both the quantum and classical cases, sits within the second nearest neighbor frequency-energy curve in Fig.~\ref{fig:f-E}. 

\noindent \textit{Conclusion} --- We have identified a previously unexplored dynamical quantum signature of particle trapped in a double-well potential: a tunneling-assisted interference effect that generates a low-frequency component in the mean particle position. It combines two major quantum phenomena, quantum interference and quantum tunneling, in one system. Unlike conventional signatures based on state reconstruction or long-time revivals, the effect appears in a first-moment observable and at the onset of the dynamics. It originates from the non-zero overlap between eigenstates situated in opposite wells and is strictly forbidden in the classical case. Given the rapid experimental progress in creating double-well traps for levitated particles, our results open a new route toward demonstrating the quantumness of the motion of non-microscopic particles.

%[Conclusion mention: quantum additional frequency enabled by tunneling (unique here), and the narrow width of the frequency that matches the classical case (more general, mention rotation) ]

% \begin{equation}\label{eq:Ry}
% R_y(\theta_0)=
%     \begin{pmatrix}
%         \cos(\theta_0/2) & -\sin(\theta_0/2)\\
%         \sin(\theta_0/2) & \cos(\theta_0/2)
%     \end{pmatrix}.
% \end{equation}
% The resulting final state is
% \begin{align}\label{eq:ANAstate}
%     \bar{\rho}_f&=(1+2p)(1-p)^2\cos^2(\theta_0/2)|\bar{\psi}\rangle\langle\bar{\psi}|\nonumber\\
%     &+(3p^2-2p^3)\cos^2(\theta_0/2)X_L|\bar{\psi}\rangle\langle\bar{\psi}|X_L\nonumber\\
%     &+(2p-3p^2+2p^3)\sin^2(\theta_0/2)Y_L|\bar{\psi}\rangle\langle\bar{\psi}|Y_L\nonumber\\
%     &+(1-2p+3p^2-2p^3)\sin^2(\theta_0/2)Z_L|\bar{\psi}\rangle\langle\bar{\psi}|Z_L,
% \end{align}

%apsrev4-2.bst 2019-01-14 (MD) hand-edited version of apsrev4-1.bst
%Control: key (0)
%Control: author (72) initials jnrlst
%Control: editor formatted (1) identically to author
%Control: production of article title (-1) disabled
%Control: page (0) single
%Control: year (1) truncated
%Control: production of eprint (0) enabled
%
 % Tell bibtex which .bib file to use (this one is some example file in TexLive's file tree)

%\appendix

%\section{More stuff...}\label{sec:more}

\end{document}